# Automated Synthesis of Assertion Monitors using Visual Specifications

Ambar A. Gadkari and S. Ramesh
Department of Computer Science and Engineering
Indian Institute of Technology Bombay, India.
ambar@cse.iitb.ac.in, ramesh@cse.iitb.ac.in

**Abstract**

*Automated synthesis of monitors from high-level properties plays a significant role in assertion-based verification. We present here a methodology to synthesize assertion monitors from visual specifications given in CESC (Clocked Event Sequence Chart). CESC is a visual language designed for specifying system level interactions involving single and multiple clock domains. It has well-defined graphical and textual syntax and formal semantics based on synchronous language paradigm enabling formal analysis of specifications. In this paper we provide an overview of CESC language with few illustrative examples. The algorithm for automated synthesis of assertion monitors from CESC specifications is described. A few examples from standard bus protocols (OCP-IP and AMBA) are presented to demonstrate the application of monitor synthesis algorithm.*

## 1 Introduction

Assertion-based verification [1-4] is gaining popularity in system level design due to its capability to combine formal methods with simulation based validation techniques. Precise and unambiguous specification is central to the success of system design and verification. Capturing high-level assertions using specification languages such as PSL/Sugar [5] or temporal logic becomes complex for interactions involving long event sequences [6]. On the other hand, manual construction of assertion monitors using native languages is error-prone and does not scale well. This paper provides a novel methodology for automated synthesis of assertion monitors for complex SoC Designs. The methodology makes use of a visual specification language called CESC (Clocked Event Sequence Chart).

CESC is a visual language designed for specifying interactions between different modules in the system with single as well as multiple clock domains typically found in SoC designs. Properties or assertions related to interactions can be captured intuitively using CESC. CESC has a precisely defined abstract textual syntax. In [7], we defined CESC and illustrated its use in the context of several real-life examples including single and multiple clock domains.

In this paper, we present an automated procedure to synthesize assertion monitors from CESC. An important feature of the procedure is that the monitor synthesized consists of a number of local monitors one for each clock domain in the given input CESC specification; the monitors communicate and synchronize with each other exchanging the information about the local states using a scoreboard-like data structure. The scoreboard dynamically maintains the information about event occurrences, which is used in implementing the causality checks within the same and across different clock domains. A formal semantics has been defined for CESC based on the synchronous language paradigm [8,9], which forms the basis for the correctness of the synthesis procedure. The methodology to automatically synthesize monitors from CESC specifications helps in smooth integration of formal specification with simulation-based verification, thus reducing the cycle time and errors involved in the manual development of monitors.

This paper is organized as follows: Section 2 provides the background and related work. Section 3 gives a quick overview of CESC language. Section 4 describes the role of monitors in assertion-based verification and formally defines the monitor. Section 5 describes the algorithm for synthesis of monitors from CESC specifications. Section 6 includes a few examples of application of the monitor synthesis algorithm on standard bus protocols. Section 7 concludes the paper.

## 2   Background

With the increasing adoption of formal verification, many property specification languages having notations similar to temporal logics or regular expressions have emerged for hardware verification. These include Acellera's Sugar [5], Synopsys' OVA [2] and Intel's FTL [10]. These languages describe various event sequences and their temporal relationships in abstract symbolic notations unfamiliar to practicing engineers. The industrial specifications often include intuitive and visual constructs such as timing diagrams, sequence diagrams, message sequence charts, which are easier to use. For example, the specifications for protocol standards such as AMBA CLI [11] and OCP-IP [12] use notations like sequence diagrams and timing diagrams to explain the cycle accurate behavior of transactions. But these notations are often informal and ambiguous. Recently, there have been efforts to formalize MSC-like notations and use them in system design [13,14]. CESC language differs from these formalisms in many ways: It offers features such as synchronizing clocks (grid lines) suited for formally specifying the interactions in hardware systems; besides, it has support for structural constructs and multiple clocks thereby making it a good choice for the system level specification of SoCs which are Globally Asynchronous Locally Synchronous (GALS) systems.

There have been efforts on formalizing other visual notations such as timing diagrams [6, 15]. But compared to CESC, these are limited in capability for specifying the complex event behaviors such as causality relationships, repetitive event sequences and multiple clocked synchronizations that are commonly found in SoC design context. Also, the structural constructs such as sequential and parallel compositions and alternative constructs, which are useful for systematic building of specifications for large system-level designs, are missing in those notations.

With assertion-based verification approaches gaining wider acceptance in industry [16, 3-5], few attempts have been made for automatic construction of monitors and protocol checkers from temporal language specifications [17,18]. They involve generation of automata equivalent to the specified property in temporal logic. The work described in this paper attempts to take this one step further by providing a methodology to synthesize assertion monitors directly from the visual specifications given in CESC.

## 3   Overview of CESC

Clocked Event Sequence Chart (CESC) language is designed for specification of the interaction behaviors in clocked systems. The simplest CESC is known as SCESC (Single Clocked Event Sequence Chart) and represents a finite duration event sequence or interaction scenario. The basic constructs provide representation for agents (referred to as instances pictured as vertical lines), synchronizing clocks (shown as horizontal grid lines) and the presence or absence of events. The events can be guarded by conditions given as propositions formed over system variables or events. Connecting arrows show the causality relationship between the events. Events occurring on external agents are shown on the frame of the chart and are referred to as environment events. Figure 1 shows an SCESC corresponding to the event sequence related to a typical read protocol within single clock domain.

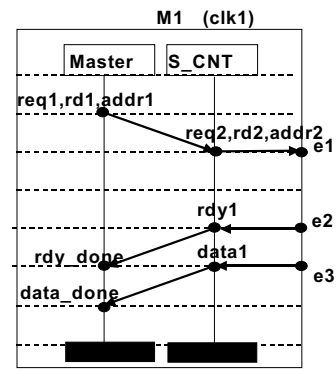

**Figure 1. Typical read protocol (single clocked)**

Various structural constructs are provided to enable hierarchical specification of complex interaction scenarios. Such constructs include sequential and parallel composition, loop, alternative, and implication. CESC constructs also include a special construct for asynchronous parallel composition to allow specification of interactions involving multiple clocks. Figure 2 shows the CESC corresponding to the read protocol involving multiple clocks.

**Semantics**

The semantics of CESC is based on clocked traces or runs. Each run describes the valuations of conditions and events along sequence of clock ticks. For formally defining the notion of a run we provide the following two definitions.

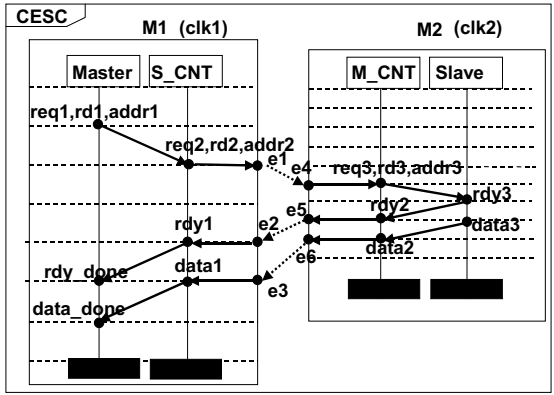

**Figure 2. Typical read protocol (multi-clocked)**

*Definition:* **State (*s*)**

A state $s \in STATES$ is defined as the assignment of truth-values to the propositions and the events. It is given as, $s = \{(f1, f2) \mid f1: PROP \rightarrow Boolean; f2: EVENTS \rightarrow Boolean\}$. The notation $\pi_i$ is used to denote the projection onto i-th element in a tuple. Thus, $\pi_1(s) = f1$ and $\pi_2(s) = f2$ for a state *s*. ♦

*Definition:* **Run (*r*)**

A run $r \in RUNS$ is defined as $r: N \rightarrow STATES$, a mapping that associates a sequence of clock ticks represented as integers with a sequence of states. $r(n)$ denotes a state at n-th clock tick. Thus, $\pi_1(r(n))$ denotes f1 and $\pi_2(r(n)) = f2$ at clock tick n. We further denote the following: $\forall p \in PROP$ (the set of all propositions), $r(n) \models p$ if $\pi_1(r(n))(p)$ = True; $r(n) \models \neg p$ if $\pi_1(r(n))(p)$ = False. Also, $\forall e \in EVENTS$ (the set of all events), $r(n) \models e$ if $\pi_2(r(n))(e)$ = True; $r(n) \models \neg e$ if $\pi_2(r(n))(e)$ = False; ♦

The semantic domain consists of a set of all possible runs. Intuitively, such a set represents an unconstrained set of behaviors that any system can exhibit. The specifications given in form of CESC constrain the permitted behaviors. Each SCESC chart specifies the constraints on the event occurrences within a finite interval on any run. The length of this interval is determined by number of grid lines denoted within a given chart. Thus, for $CLK = \{1,..., n\}$, given SCESC specifies the event occurrence pattern over 'n' clock ticks on any run. Intuitively, it can be seen (Figure 3) that for every run associated with an SCESC there is a finite interval in which the events occur according to the ordering specified by the SCESC. It may be noted that since SCESC does not provide any absolute notion of clock, the starting point of this interval is arbitrary.

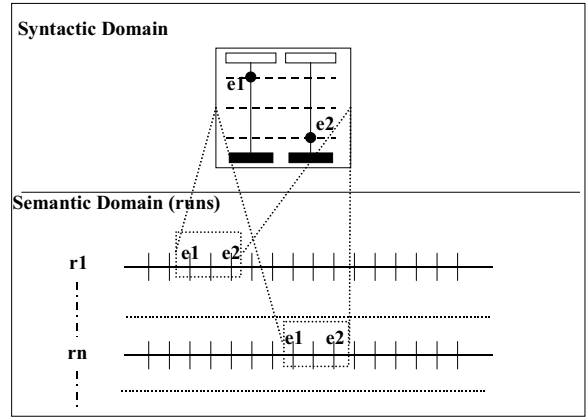

**Figure 3. Semantic mapping for SCESC**

For defining the semantics of multi-clocked CESCs a global run is defined over a global clock, which is obtained as a union of clock ticks contributed by all the component clocks in the system.

## 4  Assertion Monitors

A typical system-on-chip (SoC) verification flow is shown in Figure 4 (exclude the grey boxes).

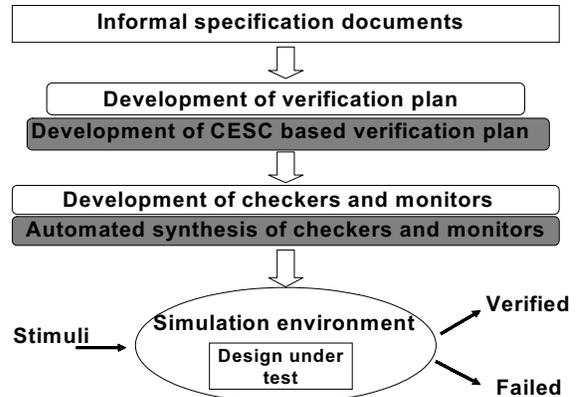

**Figure 4. SoC verification flow**

Such verification flows often involve multiple reviews and iterations during the development of verification plan and implementation of monitors / protocol checkers making the entire process time-consuming. Also the manual development of monitors introduces is prone to large number of errors. The CESC based flow suggested in [7], (refer to Figure 4 by including the grey boxes), helps in multiple ways. The verification plan consisting of different scenarios specified as CESCs is easier to develop and review and can be formally analyzed for specification inconsistencies. The automated synthesis of monitors from CESC eliminates the step of manually developing checkers and monitors, thus reducing the cycle time and making the process less prone to human errors.

The monitor corresponding to a CESC is a finite state machine that can detect the traces (runs), which exhibit the event sequence behavior specified by the given CESC. The monitor automaton is an extension of the string-matching automaton described in [19]. The monitor automaton operates on clocked event traces. Each element of the input trace is a set valuations of events and propositions given as $\{(f1, f2) \mid f1: PROP \rightarrow Boolean; f2: EVENTS \rightarrow Boolean\}$. The monitor automaton uses a dynamic 'scoreboard' for storing the information regarding the event occurrences, which is helpful in implementing the checks related to causality relationships between events during a run.

*Definition:* **Monitor**

A monitor is a finite automata defined as a 5-tuple $\langle Q, \Sigma, \delta, s_0, s_f \rangle$ where, $Q$ is a set of states. $s_0$ and $s_f$ in $Q$ represent the initial and final state respectively. $\Sigma$ is the finite input alphabet consisting of events (*EVENTS*) and propositional symbols (*PROP*). The transition function $\delta$ is a mapping from $Q \times EXP \times ACT$ to $Q$, where *EXP* is the set of logical expressions formed over *EVENTS* and *PROP* using logical connectives '$\wedge$', '$\vee$' and '$\neg$' with their standard meaning. *ACT* is a set of actions given as {Add_evt(), Del_evt(), Null}, which can be performed on the 'scoreboard'.   ♦

Following the synchronous model of systems [9], the transitions in a monitor are instantaneous and a single clock tick separates two successive transitions. The monitor automaton begins in initial state $s_0$ and reads one element of input trace in a clock step. If the monitor is in state 's' and reads element 'e' from the input trace, it takes the transition labeled 'exp/act', provided the logical expression 'exp' evaluates to True for the valuations of *EVENTS* and *PROP* given by element 'e'. The action 'act' is performed on scoreboard while taking this transition. A sequence of transitions that takes the monitor from initial state to final state is considered as the 'accepting' run and the input trace corresponding to that run represents the finite word in the language of the monitor.

An example of a monitor corresponding to an SCESC is shown in Figure 5.

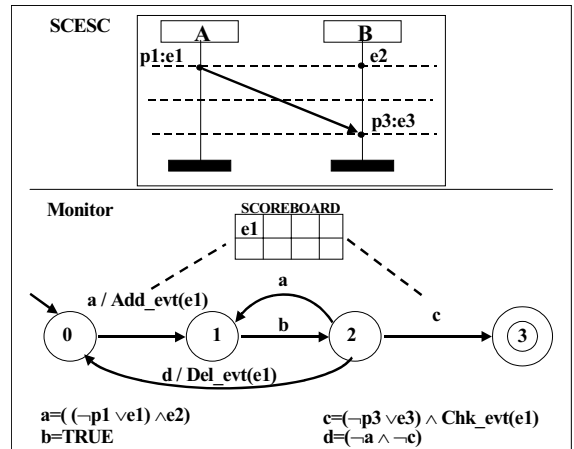

**Figure 5. SCESC and corresponding monitor**

## 5  Synthesis of Monitors

We have developed the algorithm for automated synthesis of monitors from CESC specifications. The algorithm constructs localized monitors for every SCESC, which are then combined using various composition operations. The monitors in each clock domain communicate and synchronize with the monitors in different domains exchanging the information about the local states using the dynamic scoreboard.

**Translation Algorithm -Tr**

The algorithm for construction of monitor from SCESC is described below.

```
main
Input: SCESC 'C'
Output: Monitor 'M'
begin
  Let n be the number of clock ticks in C.
  Q = {0, ..., n} /* set of states in M */
  Σ = EVENTS ∪ PROP  /* input alphabet */
  s₀ = 0, s_f = n /* initial and final states */
  P = extract_pattern(C) /* events pattern */
  δ = compute_transition_func(P, Σ)  /*
transition function */
  For every pair (e_x, e_y) in C connected by a
causality arrow do
    call add_causality_check(e_x, e_y)
  return M
end /* end of main routine */
```

The major subroutines used in main algorithm are described below.

Pattern P of events is an array where each element of the array is a logical expression formed over *EVENTS* and *PROP*. The pattern corresponding to given SCESC is extracted using following subroutine.

```
extract_pattern
input: SCESC 'C'
output: Array 'P'
begin
 For clock tick i= 0 to n in C do
 begin
  Expression 'exp' corresponding to set of
  events on each grid line in C is obtained as
  follows:
    • Event 'e' translates to exp = (e)
    • Event 'p:e' translates to exp = (¬p∨e)
    • Multiple events 'e1 ... ek' translate
      to exp = (e1∧ ... ∧ek)
  P[i] = exp
 end
 return P
end
```

The subroutine to compute the transition function for a given pattern P is described below.

```
compute_transition_func
input: P, Σ
output: δ
begin
  n = length(P)
  For s = 0 to n do
  begin
    For each valuation e ∈ 2^Σ do
    begin
      k = min(n, s+1)
      while not (P_k suffix_of T_s e) do
        k = k-1
      δ(s, e, Null) = k
    end
  end
  return δ
end
```

For describing the relation 'suffix_of' used in the above routine we first define a notion of 'matching' for the pattern. $P_k$ represents the sub-pattern of P consisting of first n elements, also referred to as 'prefix' of P of length k. Similarly, $T_s$ represents the sub-trace of input trace, after reading which the monitor is in state s. An element of pattern $e_p$ is said to be 'matched' by an element $e_T$ of input trace if the logical expression $e_p$ evaluates to True for the valuation of *EVENTS* and *PROP* given by $e_T$. A prefix $P_k$ of pattern P 'matches' with a suffix of input trace $T_s e$ (written as $P_k$ **suffix_of** $T_s e$), provided they have same length and there exists an element-by-element 'matching' between their respective elements.

The subroutine **add_causality_check($e_x$, $e_y$)** is implemented as follows: For every transition that depends on the occurrence of event $e_x$, an action 'Add_evt($e_x$)' is associated with the transition. For every transition that depends on occurrence of event $e_y$, an additional guard 'Chk_evt($e_x$)' is included along with the matching of corresponding elements from the pattern. For all the backward transitions (i.e. those taking monitor from higher numbered state to lower numbered state) all the Add_evt actions appearing on the forward path between these two states are reversed by including Del_evt actions for all the events. The example given in Figure 5 illustrates monitor automata corresponding to the given SCESC with causality arrow.

It can be shown that the finite words accepted by the monitor automaton correspond to the finite prefixes of the valid runs of the system meeting the specification given by corresponding CESC.

**Result:** Let 'M' be the monitor obtained by applying the above-described algorithm 'Tr' on SCESC 'C'. The set of runs (sequence of states) corresponding to C is given as [[C]]. Also, the language of monitor automata M is given as $\mathcal{L}(M)$. Then we can show that:

$[[C]] = \Sigma^* \bullet \mathcal{L}(M) \bullet \Sigma^\omega$.

## 6 Case Study Examples

The effectiveness of the CESC based specification and monitor synthesis approach has been studied by applying it on two standard bus protocols used widely in industry for SoC design, namely, the Open Core Protocol (OCP) [12] and Advanced Microcontroller Bus Architecture (AMBA) [11] bus protocols. We present here a few illustrative examples.

**Example 1.** Figure 6 shows the SCESC corresponding to simple read transaction (refer to page 44 of [12]) in the OCP; the corresponding monitor, which can detect the depicted scenario is also included.

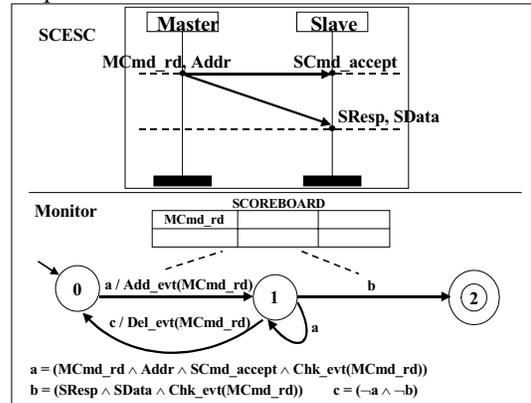

**Figure 6. OCP simple read operation**

**Example 2.** Figure 7 shows a monitor corresponding to a pipelined burst read operation (refer page 49 [12]) in OCP. Various events and related scoreboard actions are shown. (In Figure 7 NOT of action Add_evt indicates Del_evt).

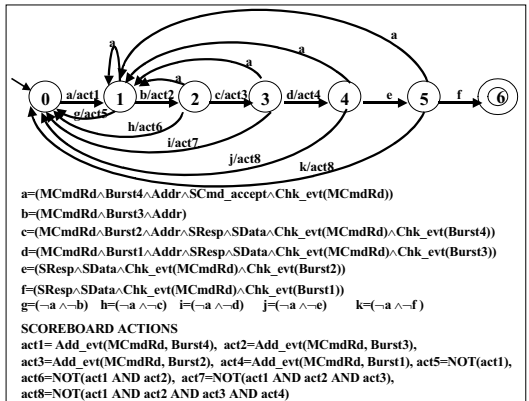

**Figure 7. OCP burst operation**

**Example 3.** We include here an example of application of CESC for specification of interactions in AMBA bus protocol. A master and bus transaction sequence (refer to page 23 [11]) is shown using SCESC with corresponding monitor in Figure 8.

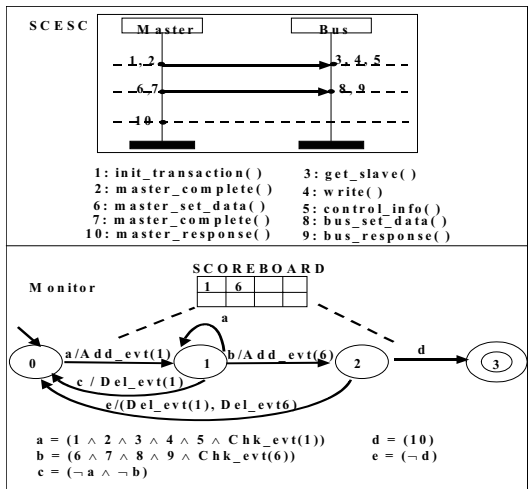

**Figure 8. AMBA AHB CLI transaction**

## 7  Conclusion

We have presented here a methodology to synthesize assertion monitors from high-level visual specifications. This effectiveness of the methodology has been tested on different parts of specifications of industrial protocols like OCP and AMBA and the initial results are very encouraging. The high level constructs of CESC were found to be very useful in structuring complex specifications and arriving at monitors in a compositional way. As part of the future work, we plan to use the synthesized monitors for checking the implementations of these protocols.

## Acknowledgements

We would like to thank Dr. Rubin A. Parekhji for discussions and valuable suggestions. We thank the Heads of Computer Science departments in IIT Bombay and IISc Bangalore for their support.